%% file: paper.tex
\documentstyle[12pt,epsfig,color]{article}
\def\baselinestretch{1.3}
\advance\textheight by \topskip
\newcommand{\comment}[1]{}

\def\beq{\begin{equation}}
\def\eeq{\end{equation}}
\def\beqn{\begin{eqnarray}}
\def\eeqn{\end{eqnarray}}

\begin{document}
 \tolerance=100000
 \input{definition.tex}
\vspace*{\fill}
\vspace{-1.5in}
\begin{flushright}
{\tt IISER/HEP/03/12}
\end{flushright}
\begin{center}
{\Large \bf Many faces of low mass
neutralino dark matter in the unconstrained MSSM, LHC data and new signals}
  \vglue 0.4cm
  Arghya Choudhury\footnote{arghyac@iiserkol.ac.in} and
  Amitava Datta\footnote{adatta@iiserkol.ac.in}
      \vglue 0.1cm
          {\it 
	  Indian Institute of Science Education and Research - Kolkata, \\
          Mohanpur Campus, PO: BCKV Campus Main Office,\\
          Nadia, West Bengal - 741252, India.\\
	  }
	  \end{center}
	  \vspace{.1cm}

\begin{abstract}

If all strongly interacting sparticles (the squarks and the gluinos) in an unconstrained 
minimal supersymmetric standard model (MSSM) are heavier than the corresponding 
mass lower limits in the minimal supergravity (mSUGRA) model, 
obtained by the current LHC experiments, then the existing data allow a variety 
of electroweak (EW) sectors with light sparticles yielding dark matter (DM) relic density 
allowed by the WMAP data. Some of the sparticles may lie just above the existing lower bounds 
from LEP and lead to many novel DM producing mechanisms not common in mSUGRA. 
This is illustrated by revisiting the above squark-gluino mass limits obtained by the 
ATLAS Collaboration, with an unconstrained EW sector with masses not correlated with the 
strong sector. Using their selection criteria and the corresponding cross section limits, 
we find at the generator level using Pythia, that the changes in the mass limits, if any, 
are by at most 10-12 $\%$ in most scenarios. In some cases, however, the relaxation of the 
gluino mass limits are larger ($\approx 20\%$). If a subset of the strongly interacting 
sparticles in an unconstrained MSSM are within the reach of the LHC, then signals
sensitive to the EW sector may be obtained. This is illustrated by 
simulating the $blj$$\etslash$, $l= e$ and $\mu$, and $b\tau j$$\etslash$ signals in 
i) the light stop scenario and ii) the light stop-gluino scenario 
with various light EW sectors allowed by the WMAP data. Some of the more general models may be 
realized with non-universal scalar and gaugino masses. 
\end{abstract}

\section{Introduction} 

The ATLAS and the CMS collaborations have been 
searching for supersymmetry (SUSY) \cite{susy} at 
the ongoing LHC 7 TeV experiments \cite{atlas0l,atlasblj,atlas1l2l,cms}. 
No signal has been seen so far. A number of phenomenological analyses
of the prospect of susy search at 7 Tev have also been published \cite{tev7}. 

The recently announced results for 1 $fb^{-1}$ data 
\cite{atlas0l,atlasblj,atlas1l2l,cms} have been presented as constraints 
in the popular minimal supergravity (mSUGRA) model \cite{msugra} \footnote {The model 
we consider is also referred to as the constrained minimal supersymmetric 
standard model (CMSSM) in the literature.}.
As expected the jets + $\etslash$ channel, 
which arises from squark-gluino pair production 
in all combinations with large cross sections and is not 
suppressed due to small branching ratios, yields the strongest 
constraints on squark-gluino masses. 

However, due to the special correlations among the superparticle 
(sparticle) masses in mSUGRA, the above mass bounds on strongly 
interacting sparticles impose stringent indirect lower bounds on the 
sparticle masses in the electroweak (EW) sector consisting of the 
sleptons and the electroweak gauginos. In many cases these model 
dependent bounds are significantly stronger than the corresponding 
direct bounds from LEP \cite{lep}. 
On the other hand, as we shall elaborate below by revising the existing limits, 
for any choice of squark-gluino masses compatible with the jets + $\etslash$ data 
there may exist a variety of EW sectors with much lighter sparticles compared to 
that in mSUGRA. In fact the jets + $\etslash$ data is only mildly sensitive to the 
electroweak sector in most cases (some exceptions will be listed 
below). Thus it is desirable to think of signals at the LHC which are 
directly sensitive to the electroweak sector.

One of the main attractive features of R-parity conserving SUSY is that 
the lightest supersymmetric particle (LSP) is stable. In many models the 
weakly interacting lightest neutralino ($\lspone$) is assumed to be the 
LSP and it turns out to be a very popular candidate for the observed 
dark matter (DM) in the universe \cite {dmrev,dmrev1}. 
The observed dark matter (DM) relic density ($\Omega h^2$) in the 
universe has been precisely measured by the Wilkinson Microwave Anisotropy Probe (WMAP)
collaboration \cite{wmap} and if 10$\%$ theoretical uncertainty is added \cite{baro} then 
DM relic density is bounded by 0.09 $\leq \Omega h^2\leq$ 0.13 at 2$\sigma$ level. 
EW sectors with relatively low mass sparticles can provide many DM producing 
mechanisms yielding relic densities consistent with the data. As mentioned 
in the last paragraph such possibilities are now mostly excluded in the mSUGRA 
framework by the LHC experiments. The negative impact of early LHC data on low mass 
neutralino DM and on the prospect of direct DM search experiments were noted in \cite{akulaprannath}. 
However, this exclusion based on strongly model dependent 
assumptions is certainly not the final verdict on  
neutralino DM since the SUSY breaking mechanism is essentially unknown. 
Hence it is worthwhile to revisit the viable DM producing 
mechanisms within frameworks more general than mSUGRA invoking as little 
model dependent assumptions as is practicable.

It has recently been emphasized \cite{arghya2} that there are many 
relic density producing mechanisms involving light electroweak 
sectors, which are practically independent of the strong sector 
leaving aside the possibility of the coannihilation of the LSP with the lighter top 
squark ($\lstop$) \cite{lspstopcoan}. Thus any model with all strongly 
interacting sparticles beyond the reach of the LHC - 7 TeV 
experiments and a relatively light electroweak sector consistent with 
the observed relic density is allowed by the LHC data. In \cite{arghya2} no 
model dependent correlation among the sparticle masses in the strong and 
EW sectors was imposed. However, for the sake of simplicity it was 
assumed that the masses in the EW sector are correlated as in mSUGRA. 
Several observable signals involving parameter spaces consistent with 
both LHC and WMAP data were proposed. In this paper we give up the last 
assumption and 
study the impact of light EW sectors in an unconstrained minimal 
supersymmetric extension of the standard model (MSSM) on both LHC 
signals and the relic density data.

Obviously the best way to test such models would be signals at a new 
$e^+ - e^-$ collider directly sensitive to only EW sparticles with masses $\gsim$ 
the corresponding lower limits from LEP. In the 
absence of such an accelerator one can look,e.g., for the clean 3l (l = 
e or $\mu$) signal from Chargino -neutralino pair production \cite {trilepton} 
and di-lepton + $\etslash$ signal from slepton pair production \cite {slepton}. 
Apriority both the signals are viable even if the strongly interacting sparticles are heavy.

The experience from the simulations of the LHC-14 TeV experiments \cite{cmstdr}, however,
does not encourage  optimistic expectation at the on going 
experiments. In the former case chargino-neutralino 
masses modestly above the LEP limits were found to be observable. 
One can improve the chargino-neutralino mass reach to some extent by considering 
2l + 1 $\tau$ and 1l + 2 $\tau$ events along with the 3l events \cite{adnabanita}. 
It is also estimated that sleptons with masses $\lsim$ 400 GeV can be probed 
at the LHC-14 TeV with $\lum$ $\gsim$ 30 $\ifb$ (see $\cite{cmstdr}$).

Models in which only a sub-set of the strongly interacting sparticles 
along with light EW sparticles are within the reach of 
the current LHC experiment are also compatible with the data. The decay of the 
strongly interacting sparticles into final states involving EW 
sparticles may provide new signals of reasonable size. Some examples 
illustrating this \cite{atlasblj,arghya2,arghya1} have already been discussed. 
However, the analyses in \cite{arghya2,arghya1} were done when either the 1 $\ifb$ data 
were not available or available in the unpublished form without many details. In 
this paper we shall improve these analyses using the published data
and more general EW sectors consistent with the WMAP data as elaborated above. 
Our main goals are to check the impact of the more general models on i) the squark-gluino mass
limits already obtained within the framework of mSUGRA and 
ii) the viability of the signals proposed in \cite{arghya2,arghya1} for more general scenarios.

In the simplest scenario considered in this paper following 
\cite{arghya2,arghya1}, the lighter stop squark $\lstop$ is assumed to 
be the only strongly interacting sparticle accessible to the current LHC 
experiments. Other groups have also investigated the light $\lstop$ 
scenarios \cite{sundrum,bmnishita}. If this squark is the next lightest supersymmetric particle 
(NLSP), then it will decay via the loop induced final state consisting 
of a charm quark and the LSP. As noted earlier the DM relic density may 
be produced in such a scenario via $\lstop$-LSP coannihilation. This 
channel has been investigated recently but does 
not appear to be very promising \cite{huitu,biyanyin}. Recently novel signals of 
$\lstop$ NLSP have been proposed \cite{dreeslightstop,tongli}. 
However, the competition between the above mode and the four body decay of 
the $\lstop$ NLSP \cite{djouadi} may further complicate the issue. The consequences of this competition in the 
context of the Tevatron experiments have already been discussed \cite{sibu_stop}. 

It was recently pointed out in ref \cite{arghya2,arghya1} that if 
this squark decays dominantly into the lighter chargino 
($\charginopm$) and a b-quark,  then  viable signals 
sensitive to the EW sector may appear in the $blj$$\etslash$ and 
$b\tau j$$\etslash$ channels. However, for $\lstop$ 
within the reach of the ongoing LHC experiments the above signals cannot 
survive the strong cuts on $\etslash$ and $M_{eff}$ usually employed for 
general squark-gluino searches in jets + $\etslash$ by the LHC collaborations. 
Softer dedicated cuts need to be employed for this signal as shown in \cite{arghya2,arghya1}. 
In this paper we shall revisit the 
signals for  more general EW sectors as elaborated above.
 
Another class of models interesting in the context of the LHC 
experiments are the ones with both the $\lstop$ and the gluinos are within the 
reach of the current experiments: the light stop-gluino (LSG) model  \cite{arghya2,bmnishita}. 
This scenario can be realized in models with non-universal scalar and 
gaugino masses at the GUT scale \cite{arghya2}. The 
same signals as discussed in the last paragraph may arise in this case 
also via the production of gluino pairs which then decay dominantly into 
t$\lstop$ pairs. It may be recalled that the ATLAS collaboration has 
already investigated the $blj$$\etslash$ signal \cite{atlasblj}. However, the hard cuts 
proposed for separating the background eliminates the events from the direct $\lstop$ 
pair production. As a result the gluino mass limit (500-520 GeV) obtained 
by them is by and large independent of the $\mlstop$. On the other hand the 
alternative selection criteria with softer cuts as proposed in \cite{arghya2,arghya1} can separate 
the $blj$$\etslash$ events from $\lstop$ pair production from the ones coming from 
gluino pair production. In this paper we shall follow this approach and shall revisit the signal for the more 
general EW sectors discussed above. Moreover the complementary channel $b\tau j$$\etslash$ 
which is more important when the electroweak gauginos decay dominantly in $\tau$-rich 
final states was not considered in \cite{atlasblj}. This signal was studied 
for the first time in \cite{arghya2}.

The plan of the paper is as follows. Section 2 contains all the results obtained 
in this paper and the related discussions. In Section 2.1 we list the input parameters 
for the unconstrained EW sector. The main departures from the mSUGRA spectrum are duly emphasized. 
The possibility of accommodating these departures with non-universal scalar and gaugino masses 
are discussed as and when appropriate. In sub-section 2.2 the parameter space in the 
unconstrained EW sector consistent with WMAP data is delineated. DM relic density 
producing mechanisms which are not allowed in mSUGRA but occur frequently in more general 
scenario are pointed out. Revision of the squark-gluino mass bounds in mSUGRA for more general EW sectors 
consistent with WMAP data is presented in the sub-section 2.3. Novel signal in the 
$blj$$\etslash$ and $b\tau j$$\etslash$ channels in the light stop and LSG scenarios 
are also included in this section. Our conclusions will be summarized in Section 3.

\section{Results and Discussions :}

\subsection{Models and Parameters : }

The popular mSUGRA model \cite{msugra} has only five free parameters including soft SUSY breaking terms. 
These are $m_0$ (the common scalar mass), $\mhalf$ (the common gaugino mass), $A_0$ (the common 
trilinear coupling), all given at the gauge coupling unification scale ($M_G$ $\sim  2 \times 10^{16}$ GeV); 
the ratio of the Higgs vacuum expectation values at the electroweak scale namely tan$\beta$ and 
the sign of $\mu$. The magnitude of $\mu$ is determined by the radiative electroweak symmetry breaking condition.

In contrast in the phenomenological MSSM $M_1, M_2, M_3$ are the three gaugino mass 
parameters at the weak scale and no special relation among them is 
assumed. It is known for a long time that even in mGUGRA type models
where masses at the weak scale are determined by some boundary 
conditions at a high scale, a variety of gaugino mass relations at the weak 
scale may emerge due to non-universal boundary conditions for gaugino 
masses at $M_G$ \cite{nonunigauge}.

We also consider the following weak scale parameters. $M_{\tilde 
q_i}$ is the mass parameter for the ith (i = 1, 2 and 3 is the 
generation index) generation squarks of the L-type belonging to a 
doublet of $SU(2)_L$.  $M_{\tilde u_i}$ ($M_{\tilde d_i}$) is the mass 
parameter for the R-type singlet up (down) squarks. Similarly $M_{\tilde 
l_i}$ ($M_{\tilde r_i}$) is the left (right) type slepton mass parameter 
for the ith generation. $A_t$, $A_b$, $A_{\tau}$ are the 3rd generation 
trilinear soft couplings. The other input parameters are the mass of 
pseudoscalar Higgs boson $m_A$, the higgsino mixing parameter $\mu$ and 
tan$\beta$.

In our numerical analysis we have assumed that all squarks and gluinos 
are beyond the reach of the LHC 7 TeV run \footnote{ All masses and 
parameters having the dimension of mass are in GeV unless stated 
otherwise. }. For the sake of simplicity we have taken $M_{\tilde q_1}$ 
= $M_{\tilde q_2}$ = $M_{\tilde q_3}$ = $M_{\tilde u_1}$ = $M_{\tilde 
u_2}$ = $M_{\tilde u_3}$ = $M_{\tilde d_1}$ = $M_{\tilde d_2}$ = 
$M_{\tilde d_3}$ = 1.5 TeV. For the relic density computation in this 
subsection we also take  $M_3$ (which defines the gluino mass) = 1.5 TeV at the weak scale. 

We have further assumed that $M_{\tilde l_1}$ = $M_{\tilde l_2}$ = 
$M_{\tilde r_1}$ = $M_{\tilde r_2}$ = $M_{\tilde l}$ ; $M_{\tilde l_3}$ = $M_{\tilde 
r_3}$; but these common masses at the weak scale are treated as free 
parameters. The equality of the slepton mass parameters at the 
weak scale is not realized in mSUGRA, where L-type sleptons are typically 
heavier than the R-type. Since the sneutrino mass is correlated
with that of L-type sleptons, it also exceed R- type slepton masses
over most of the parameter space. On the other hand models with 
non-universal scalar masses at the high scale can yield considerably 
different mass hierarchies among the two types of sleptons at the weak 
scale.The running of the common scalar masses at the SUSY breaking scale 
(say, the Planck scale) to  the GUT scale $(M_G)$  
may create the above nonuniversality when L and R type sleptons belong to 
different representations of the GUT group \cite{polonsky}. 
When a GUT group breaks down to a group of lower rank, certain 
$U(1)$ symmetry breaking D-terms can also lead to such 
non-universality \cite{dterm}.

Relatively light sleptons and, consequently, sneutrinos can affect the 
DM relic density production through LSP annihilation via slepton exchange
and / or via LSP-sneutrino co-annihilation in a large region of the parameter 
space where the sneutrino happens to be the next lightest supersymmetric 
particle (NLSP). In the context of collider phenomenology, the relative 
probabilities of m-lepton + n-jet + $\etslash$ signatures, for different m and n, 
may turn out to be quite distinct in comparison to mSUGRA 
\cite{nonuniscapheno} in such models. This may distinguish between different models.

In addition we assume $A_t$ = $A_b$ = $A_{\tau}$ = -600;
$\mu$ = 362.0 ; $m_A$ = 1 TeV and tan$\beta$ = 10.
Moreover since our focus is on a light EW sector the mass parameters 
in this sector are restricted - somewhat arbitrarily - to be less than 
200. As a result we have either a bino dominated LSP or an admixture 
of bino and wino with negligible Higgsino component. 

In this parameter space the computed mass of the lighter Higgs scalar ($m_h$), 
after taking into account a theoretical uncertainty of 3 \cite{mhcorrection}, satisfies 
the LEP bound $m_h$ $\ge$ 114.4 \cite{lephiggs}. Further discussion of 
$m_h$ in the light of more recent experiments is given below.

We have used micrOMEGAs (v.2.4.1) \cite{micromegas} for computing the 
DM relic density. The susy particles spectra and the decay branching 
ratios (BRs) have been computed by SUSPECT \cite{suspect} and
SDECAY \cite{sdecay}.

In Fig. 1 we plot $\Omega h^2$ vs $M_1$ for different choices of the other 
parameters in the EW sector. We have also checked that varying the $\mlstop$ 
or $\mgl$ does not affect the DM relic density, unless the difference between 
$\mlstop$ and $\mlspone$ is so small that the possibility of stop-neutralino 
coannihilation \cite{lspstopcoan} opens up. 

To begin with we have searched for the parameter space allowed by the 
WMAP data. Results are given in Fig. 1. Here we have plotted $\Omega h^2$ vs $M_1$ 
for different $M_2$ and common slepton masses. We have chosen the 
following five representative sets of $M_2$ and the common slepton mass of the third 
generation ($M_{\tilde l_3}$ = $M_{\tilde r_3}$): model-1 (110,115), 
model-2 (110,150) , model-3 (150,115), model-4 (150,200) and 
model-5 (200,200). The choices $M_2$ = 110, 150 and 200 yield 
$\mcharginopm$ = 106 ,145 and 191 respectively while $M_{\tilde l_3}$ = 
$M_{\tilde r_3}$ = 115, 150 and 200 result in $\mstauone$ = 89, 131, 186 
respectively. For each combination of $M_2$ and $M_{\tilde r_3}$, we 
have varied the common slepton mass ($M_{\tilde l}$) for the first two
generations  in steps of 10 in the range 100 to 200. 
Thus we allow the lighter $\stau$ mass eigenstate ($\stauone$) to be lighter 
or heavier than the sleptons belonging to the first two generations. This is also 
a significant departure from the mSUGRA spectrum. It should be noted that we 
have several choices where the masses of the sparticles in the 
electroweak sector are just above the corresponding LEP limits. 


\begin{figure}[!htb]
\begin{center}
\includegraphics[angle =270, width=1.0\textwidth]{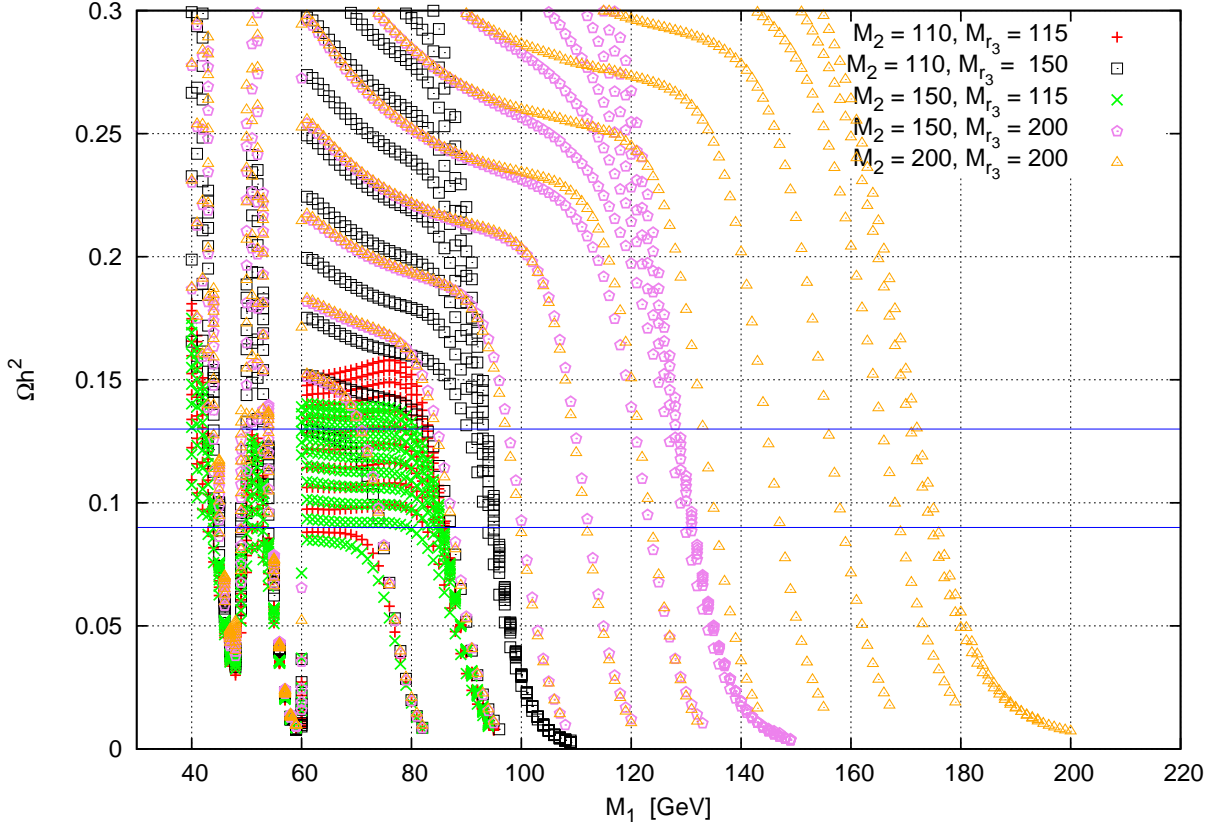}
\end{center}
\caption{ {\footnotesize $M_1$ vs $\Omega h^2$ for different choices of $M_2$, $M_{\tilde r_3}$ (= $M_{\tilde l_3}$) 
and $M_{\tilde l}$ (= $M_{\tilde l_1}$ = $M_{\tilde l_2}$  = $M_{\tilde r_1}$ = $M_{\tilde r_2}$). Blue horizontal lines represent 
the WMAP allowed DM relic density band (0.09 to 0.13).   }}
\end{figure}

In Fig 1, each color or point shape corresponds to a particular choice of 
$M_2$ and $M_{\tilde r_3}$ as indicated in the upper right corner.
The ten different lines of the same color or type in Fig. 1 correspond to 
different choice of $M_{\tilde l}$. 

Throughout this work the pole mass of the top quark (running bottom 
quark mass evaluated in the $\overline{MS}$ scheme) will be 
taken as $m_t$ $(m_b)=$ 173.2 (4.25). We have also included the current 
constraints on squarks, sleptons, gluino, chargino, neutralinos and the 
lighter scalar Higgs boson masses obtained from collider data. For 
example in the CMSSM models with tan$\beta$ = 10, $A_0$ = 0 and $\mu$ 
$>$ 0, ATLAS has excluded squarks and gluinos having equal mass below 
950 GeV \cite {atlas0l}. We shall assume that $\mcharginopm \ge $ 103.5, 
all the sleptons (except the lighter stau mass eigenstate) are heavier 
than 100. This is basically a simplified form of the LEP limits. The 
lighter $\stau$ mass eigenstate is assumed to be heavier than 86 as 
required by the LEP data. We emphasize that there is no model 
independent bound on $\mlspone$ from collider data. From the 
chargino mass bound from LEP there is a bound $\mlspone$ $>$ 50.8 for 
tan$\beta$ = 10 and $\mu$ $>$ 0, which is valid in mSUGRA. From LHC, 
WMAP and XENON100 \cite{xenon100} there is a bound $\mlspone$ $\ge$ 160 in mSUGRA \cite 
{boer}. We have only considered parameter spaces resulting in 114.4 \cite{lephiggs} $< m_h < 
127$ \cite{higgsatlas,higgscms}. We have checked that recent hint of a peak at 
$m_h \approx 125$ reported by the LHC collaborations \cite{higgsatlas}, which 
is not yet statistically significant, can be accommodated if we make 
$A_0$ large (say 1.5 TeV). However, our parameter space allowed by the 
relic density data is not drastically affected by the choice of $A_0$.
The impact of the $m_h$ bound on different SUSY models and various 
observables has been considered by several groups \cite{higgs125}.


\subsection{Summary of Fig.1 : }

There are five sets of data :

\begin{itemize}

\item {\bf Model- 1 }: The red points (marked by +) represent the computed 
$\Omega h^2$ in this parameter space. Here both $\chonepm$ and $\wt \tau_1$ 
lie just above the LEP limit. The latter is the NLSP for the entire
parameter space scanned and its mass sets the limit for the LSP mass.

\item {\bf Model-2 }: The squares represent the computed relic 
density. Here the mass of the $\chonepm$ is just above the LEP limit  and 
it is the NLSP for the whole parameter space studied. The mass of $\wt 
\tau_1$ is somewhat away from the LEP limit.

\item {\bf Model-3 }: The green cross marks represent the parameter 
space with $M_2$ = 150, $M_{\tilde r_3}$ = 115.  Here the mass of $\wt 
\tau_1$ is just above the LEP limit and it is the NLSP in our study. 
However, the mass of $\chonepm$ is away from the LEP  limit.

\item {\bf Model-4 }: Relic densities are represented by the violet pentagons. 
In this case both $\chonepm$ and $\wt \tau_1$ have masses far away from the 
the LEP limit. For relatively low common slepton masses the sneutrino 
is the NLSP.
Otherwise the chargino is the NLSP. 

\item {\bf Model-5 }: In this case the orange triangles represent the 
relic densities. Over most of the parameter space the sneutrino is the NLSP. 
For the highest common slepton mass (200), the lighter stau, the 
sneutrino and the chargino have closely spaced masses.
\end{itemize}

We have also demarcated the WMAP allowed DM relic density band (0.09 to 
0.13) by the blue horizontal lines. In the next section we will take different 
benchmark points from this region for studying the impact of this more general EW 
sector on LHC physics.

The importance of LSP annihilation via relatively low mass R-type slepton 
exchange (Bulk annihilation) in producing the observed relic density is 
well known in mSUGRA \cite {dmrev,dmrev1}. The corresponding region of 
the parameter space (the Bulk region) is, however, strongly disfavoured 
by the current LHC constraints as they do not permit the sleptons to be 
sufficiently light. In the unconstrained MSSM this mechanism retains it 
importance for certain ranges of the LSP mass where both L and R type slepton 
exchange may be important. However, even for smaller and larger LSP masses there 
are many other potentially important relic density producing mechanisms.

In order to streamline the discussions, we divide the whole region 
allowed by the WMAP data into three parts according to the value of 
$M_1$ as follows :

{\bf I)} 40 $\lsim$ $M_1$ $\lsim$ 60 : In this region different choices 
of $M_2$ consistent with LEP data give almost the same DM relic density, 
which, however, depends strongly on the slepton masses. The neutralinos 
studied in this paper are either bino dominated or mixture of bino and wino. 
For such neutralinos, $\mlspone$ significantly smaller than 40 are not 
allowed due to the LEP lower bounds on chargino and slepton masses and 
the WMAP constraints. In fact we have checked that $\mlspone$ $\gsim$ 35 
in the parameter space we have scanned.

Three main mechanisms or combinations operate in this case: 
i) bulk annihilation,
ii) annihilation mediated by a Z boson of low virtuality or 
iii) mediated by the lighter Higgs scalar (h) of low virtuality.
It should be noted that even if we ignore the indirect bound on $\mlspone$ from 
the LHC bounds in mSUGRA, a large part of this region is strongly disfavoured
in mSUGRA due to the LSP mass bound from LEP inferred from the chargino mass 
limit. We also stress that no co-annihilation process can play any role in this 
parameter space due to the LEP lower bounds on EW sparticle masses.

For stau mass a little above the LEP bound (Model 1 or 3), $M_{\tilde l}$ 
not very far from the corresponding bound 
and the LSP mass below the Z resonance, this happens through pure 
bulk annihilation for a range of slepton masses(e.g., with $M_1$ = 
40, $M_{\tilde l}$ $ < $130). For higher $M_1$ on both sides of the Z 
resonance, both bulk annihilation and annihilation via a virtual Z contributes adequately 
in all models. However, the slepton masses for which this happens is 
model dependent.  For  $M_1$ on the Z resonance or in its immediate vicinity 
the annihilation cross section is too large and $\Omega h^2$ is too
small for all LEP allowed slepton and stau masses.  
 
Well above the Z resonance,  the Z contribution begins to reduce.
Now process  iii) takes over. Combination of i) and iii) 
is now the potential mechanism. This happens in all models 
1 - 5 but the range of $M_{\tilde l}$ is again model dependent. There is  
even a small parameter space where i), ii) and iii) all contributes 
significantly. 

On the h resonance or in its immediate neighbourhood
$\Omega h^2$ is again too small. 
At $M_1$ = 56, $\Omega h^2$ falls abruptly and nearly at $M_1$ = 59 we see a 
minima. This is due to the higgs resonance (here computed $m_h$ = 113 $\pm$ 3 (theoretical
uncertainty). Also here the relative contribution to 1/($\Omega h^2$) is largest 
from the process $\lspone \lspone$ $\ra$ $b$ $\bar b$ (nearly 80 $\%$).

{\bf II)} 60 $\lsim$ $M_1$ $\lsim$ 80 : A large number of points in 
model 1 and 3, with $\stauone$ NLSP, for different common slepton masses 
lie in 
the region  allowed 
by WMAP data. The value of $\Omega h^2$ remains almost constant for LSP masses in 
this range for fixed values of other parameters. Pure bulk 
annihilation is the source of DM relic density. 

In models 2, 4 and 5 with heavier $\stauone$, pure bulk annihilation is 
not viable. However, if the sneutrinos belonging to the first two 
generations are the NLSP, then a small but significant contribution from 
LSP- sneutrino coannihilation along with bulk annihilation may produce 
the required relic density. For each $M_1$ there is a narrow range of 
sneutrino masses where this happens. This mechanism is indicated by a 
sharp fall of the relic density over a narrow range of $M_1$.

{\bf III) 80 $\lsim$ $M_1$ $\lsim$ 200 :}
In models 1 and 3, the LSP-$\stauone$ coannihilation becomes effective 
for a small range of $M_1$ just above 80.

In models 2, 4 and 5 with heavier $\stauone$, bulk annihilation alone can 
not produce the observed relic density. If the 
lighter chargino is the NLSP, several choices of the common 
slepton mass
yield approximately the same relic density (see, e.g., the squares
(model 2) in the neighbourhood of $M_1 \approx 95$). In fact $\Omega h^2$ in this case
primarily depends on the LSP and 
chargino properties alone. LSP annihilation into W pairs and LSP-Gaugino 
co-annihilation are the significant mechanisms for relic density
production. 
If, on the other hand, the sneutrino is the NLSP,  
then sneutrino co-annihilation and bulk annihilation may serve the purpose.
This happens for a small range of the common slepton mass for each $M_1$. 
Several examples are shown in Fig. 1.

\subsection{Collider Signals  }

In this section we discuss the collider signatures corresponding to some  
benchmark points allowed by both WMAP (see Section 2.2) and LHC data.

In this paper all leading order (LO) signal cross-sections have been 
computed by CalcHEP \cite{calchep} unless otherwise stated. For any two 
body final state (except for QCD processes) with identical particles or 
sparticles both the renormalization and the factorization scales are 
taken as, $\mu_R = \mu_F = M$, where $M$ is the mass of the particle or 
sparticle concerned \footnote{See footnote 4.}. For two unequal masses 
in the final state the scales are taken to be the average of the two. 
For QCD events the scales have been chosen to be 
equal to $\sqrt {\hat s}$ which is the energy in the parton CM frame, 
and the cross-section is computed by Pythia \cite{pythia}. All LO cross-sections are 
computed using CTEQ5L parton density functions (PDFs) \cite{cteql}.

The next to leading order (NLO) cross-sections for the signal processes 
have been computed by PROSPINO \cite{prospino} using the CTEQ5M PDFs.
The K-factors are computed by comparing with the LO cross-section. The LO 
cross-sections from PROSPINO agree well with  CalcHEP for the same choice of 
the scales. The NLO cross sections are used to compute revised mass limits and 
the signal sizes presented in this section.

The NLO background cross-sections are not known for some 
backgrounds - in particular for the QCD processes.  
For computing the significance of the signal we conservatively multiply 
the total  LO background by an overall factor of two. 

We have considered the backgrounds from  $t \bar t$, QCD events ,
$W$ + $n$-$jets$ events and $Z$ + $n$-$jets$, where $W$ and $Z$  decays into all channels.
After the final cuts  $t \bar t$, $W + 1j$ and $W + 2j$ are the main backgrounds. 
$t \bar t$ events are generated using Pythia and the LO 
cross-section has been taken from CalcHEP which is 85.5 pb.
QCD processes are generated by Pythia in different $\hat p_T$ bins :
$25 \le \hat p_T \le 400$, $400 \le \hat p_T \le 1000$ and 
$1000 \le \hat p_T \le 2000$ , where $\hat p_T$ is defined in
the rest frame of the  parton collision.
The main contribution comes from the low $\hat p_T$ bin,
which has a cross-section of $\sim 7.7 E+07$ pb.
However, for the other two bins the background events are
negligible.

For $W$ + $n$-$jets$ events we have generated events with $n=0,1$ and 
$2$ at the parton level using ALPGEN (v 2.13) \cite{alpgen}.
We have generated these events subjected to the condition
that $P_T^j > 20$, $\Delta R(j,j) \ge 0.3$ and
$\vert \eta \vert \le 4.5$. These partonic events have been fed to 
Pythia for parton showering, hadronization, fragmentation and decays etc.

We have used the toy calorimeter simulation (PYCELL) provided in Pythia
with the settings described in \cite{arghya2,arghya1}. Lepton $(l=e,\mu)$ selection, 
$b$- jet tagging and $\tau$- jet identification are also implemented 
following these references. 


The strongest limits on squark, gluino masses come from the jets + $\etslash$  channel. 
The ATLAS  group has introduced five sets of selection criteria (SC) (see Table 1 of \cite{atlas0l}) 
for new physics search in this channel. 
The observed number of events in this channel and the SM background estimated from the data 
for each of the above SC lead to a model independent upper bound on $\sigma_{new}$, 
the effective cross-sections for any new physics scenario. 
Five such bounds, thus obtained, are  22 fb, 25 fb, 429 fb, 27 fb and 17 fb 
respectively at 95$\%$ confidence level. These bounds can be used to obtain exclusion 
contours in any specific SUSY breaking model.

Some examples of the limits (L1-L4) obtained by the ATLAS collaboration  
in the mSUGRA/CMSSM model (See Fig. 2 (right) of \cite {atlas0l}) 
for tan$\beta$ = 10, $A_0$ = 0 and $\mu$ $>$ 0 are given below.  

L1)Squarks and gluinos of equal mass are excluded for 
masses below 950 GeV. Here  the average squark mass is 
considered.

From the exclusion plot 
of ATLAS  it also follows that $m_0$ = 
1.0 (1.5) TeV and $\mhalf$ $\sim$ 300 (210) lies on the exclusion 
contour. It then follows that 

L2) For  average squarks mass of 1.1 TeV gluino masses below 780 GeV
are excluded.

L3) For  average squarks mass of 1.5 TeV gluino masses below 600 GeV
are excluded.

For the validation our simulation, we focus on the above exclusion contour. 
By definition $\sigma_{new}$ = $\sigma$A$\epsilon$, where $\sigma$ is the raw production 
cross-section in any new physics scenario, A is the acceptance and 
$\epsilon$ is the efficiency \cite{atlas0l}. 
We have chosen several points on this exclusion contour. For 
each point we compute A and $\epsilon$ corresponding to all the five SC of ATLAS from our simulation. 
For each mSUGRA point $\sigma$ in the NLO is computed by PROSPINO. We find that 
each computed $\sigma_{new}$ is reasonably close to at least one of the upper 
limits (see above) obtained by ATLAS. This reflects that our simulation in mSUGRA would 
lead to an exclusion contour pretty close to the one obtained by ATLAS. We follow this procedure 
to obtain new limits in other models. 

We now turn our attention to more general models where the masses of the sparticles 
in the strong and EW sectors are assumed to be uncorrelated. We consider various 
EW sectors allowed by the WMAP data as illustrated in Fig. 1  of this paper
and compute the squark-gluino mass limits from ATLAS  bounds on $\sigma_{new}$.

\begin{table}[!htb]
\begin{center}
\begin{tabular}{|c|c|c|c|c||c|c|c|c|}
\hline
 Points &$M_1$&$M_2$&$M_{\tilde l}$	&$M_{\tilde l_3}$	&$\mlspone$ 	&$\mchonepm$	&$\mstauone$	& $m_{\wt l_L} $	\\
	 &	&	&		&			&		&		& 		&			\\
\hline
P1	 &45	&150	&160		&200			&43		&145		&186 		&166			\\
\hline
P2	 &128	&150	&160		&200			&121		&145		&186 		&166			\\
\hline
P3	 &43	&150	&200		&115			&40		&145		&89 		&205			\\
\hline
P4	 &81	&150	&200		&115			&77		&145		&89 		&205			\\
\hline
P5	 &45	&110	&150		&150			&42		&106		& 131		&157			\\
\hline
P6	 &99	&150	&120		&200			&93		&145		& 186		&128			\\
\hline
P7	 &94	&110	&200		&150			&88		&106		& 131		&205			\\
\hline
       \end{tabular}
       \end{center}
          \caption{ {\small Parameters and mass spectra corresponding to different benchmark points
taken from different regions of Fig. 1. }}
          \end{table}

For a better understanding of the revised limits, we also select some 
benchmark EW sectors from Fig.1 (see Table 1). The corresponding BRs
relevant for this discussion are presented in Table 2. Some important 
characteristics of the selected points are given below.

\begin{itemize}

\item The point P1 corresponds to a low $\mlspone$ not allowed by the LEP data in mSUGRA. 
\item For the point P2, the  BR of the mode $\chonepm \ra \lspone l \nu_{l} $, $l=e$ and $\mu$, is large 
(78 $\%$) compared to what is obtained over most of the mSUGRA parameter 
space. This is due to the fact that the decay $\chonepm \ra \lspone W $ is not allowed 
and the common slepton mass  is relatively low. 
\item For  P3 and P4, the combined  BR of the modes 
$\chonepm \ra\stau_{1}\nu_{\tau}$ and  $\snutau \tau $ is 100 $\%$. 
\item In examples P5 and P7,  $\chonepm$ is the NLSP. 
But the mass difference ($\Delta m$) between the $\lspone$ 
and $\chonepm$ is large  in P5 (65)and much smaller in P7 (19). The BR 
of the mode $\chonepm \ra \lspone l \nu_{l} $ is 48 (23) $\%$ for the 
point P5 (P7). 
\item In P6, $\snul$ is the NLSP, $\chonepm$ is lighter than $\stauone$ 
but heavier than $\wt{l}_L$ and $\snul$. Here BR of $\chonepm \ra \snul l $ 
is 87$\%$ and $\chonepm \ra \wt{l}_L \nu_{l}$ is 13 $\%$.
\end{itemize}

\begin{table}[!htb]
\begin{center}
\hspace{-2 cm}
\begin{tabular}{|c|c|c|c|c|c|c|c|}
\hline
Decay Modes				&P1 	&P2	&P3	&P4	&P5	 &P6	&P7	\\
\hline



$\chonepm \ra \lspone q q\prime $	&-	&17	&-	&-	&30	 &-	&51	\\
$\quad    \ra \lspone l \nu_{l} $	&-	&78	&-	&-	&48	 &-	&23	\\
$\quad    \ra \lspone \tau \nu_{\tau} $	&-	&5	&-	&-	&22	 &-	&26	\\
$\quad    \ra\stau_{1}\nu_{\tau}$	&-	&-	&36	&36	&-	 &-	&-	\\
$\quad    \ra \lspone W        $	&100	&-	&-	&-	&-	 &-	&-	\\
$\quad    \ra\snutau \tau	$	&-	&-	&64	&64	&-	 &-	&-	\\
$\quad    \ra\snul l	$		&-	&-	&-	&-	&-	 &87	&-	\\
$\quad    \ra \wt{l}_L \nu_{l}$		&-	&-	&-	&-	&-	 &13	&-	\\

\hline\hline
$\stau_{1} \ra \lspone \tau 	$	&75	&49	&100	&100	&78	 &52	&53	\\
$\quad    \ra  \lsptwo \tau 	$	&9	&21	&-	&-	&8	 &19	&19	\\
$\quad    \ra \chonepm \nu_{\tau} $	&16	&30	&-	&-	&14	 &29	&28	\\
\hline
       \end{tabular}
       \end{center}
          \caption{ {\small The BRs ($\%$) of the dominant decay modes of $\chonepm$ and $\stau_{1}$  for the benchmark points.} }
          \end{table}
In order to compare the new limits with the ones in mSUGRA, we fix the average 
squarks mass as in the examples L1, L2, L3 given above and vary $\mgl$.
We then compute the jets + $\etslash$ signature 
corresponding to EW sectors randomly chosen from the points in Fig. 1 allowed 
by WMAP data. 
If a gluino mass yields  cross sections smaller than the  
ATLAS upper bounds on $\sigma_{new}$ for all selection criteria, we reduce it keeping 
average squark mass fixed. In this way we obtain a relaxed limit on $\mgl$ for a fixed 
average squark mass when the cross-section exceeds the upper bound on 
$\sigma_{new}$ \cite{atlas0l} for at least one selection criteria. 

For most  of the points considered by us, the $\mgl$ limit changes at most by 10 
to 12 $\%$ compared to the mSUGRA limit. For e.g., for the benchmark point P1 the 
lower bounds on $\mgl$ in L1, L2, L3 are reduced by 50 to 70. 
As noted earlier this is due to relatively low $\mlspone$.

The leptonic BR of the $\chonepm$ is large for 
P6. The lepton veto in the ATLAS selection criteria relaxes 
the limits on $\mgl$ in all cases by 70-80. This
is reminiscent of the relaxation of the $\msq$ - $\mgl$ mass limits 
obtained from Tevatron data (see the 1st paper of \cite{nonuniscapheno}) 
for similar reasons.  

On the other hand if BR($\chonepm \ra \snutau \tau ,\stauone \nu_{\tau}$) 
= 100 $\%$ then the limit on $\mgl$ is strengthened by 5-10 $\%$ due to the 
large BR of hadronic $\tau$ decays. The relative changes in the limits noted by us 
are typically of the order of the uncertainties in the production cross sections 
due to the choice of QCD scale and/or the parton density function. 
This simple exercise illustrates that ATLAS limits on  squarks-gluino masses 
are approximately valid for many different electroweak sectors quite different from that in mSUGRA. 
However, we also arrive at the important conclusion that slepton and / or chargino masses 
just above the corresponding LEP limits are very much allowed by the current LHC data. 

\begin{table}[!htb]
\begin{center}
\begin{tabular}{|c||c||c|}
\hline
ATLAS CUTS \cite{atlas0l}& mSUGRA point with		&Benchmark point P1   	\\
`High Mass' 		&$m_0=2500,\mhalf=185 $		& with $\mgl$ = 550	\\
			&$A_0=0$ and $tan\beta$=10	&			\\
\hline
$P_{T}$ of leading jet $\ge$ 130	&92 	$\%$	& 76	$\%$		\\
\hline
$\etslash \ge$ 130 			&54 	$\%$	& 33 	$\%$		\\
\hline
$P_{T}$ of 2nd jet $\ge$ 80 		&75 	$\%$	& 68	$\%$		\\
\hline
$P_{T}$ of 3rd jet $\ge$ 80 		&79	$\%$	& 78	$\%$		\\
\hline
$P_{T}$ of 4th jet $\ge$ 80		&57	$\%$	& 61	$\%$		\\
\hline
$\etslash / M_{eff} \ge$ 0.2 		&47	$\%$	& 32	$\%$		\\
\hline
$M_{eff} \ge$	1100 			&29	$\%$	& 18	$\%$		\\
\hline

       \end{tabular}
       \end{center}
          \caption{ {\small Efficiency table for different cuts employed by ATLAS 
(see Table 1 of \cite{atlas0l}) for deriving in mSUGRA the limit L4 (see text) and 
the corresponding efficiencies for the benchmark point P1. }}
          \end{table}
L4) For heavy squarks (say 2.5 or 3.0 TeV) ATLAS  data has excluded 
gluinos with masses in the range 550 to 570 . 

Now keeping the average squarks mass as in L4 but choosing the EW sectors from 
Figure 1, we have checked the change in the limit on gluino mass. 
Some interesting examples are given below. Models with EW sector as 
in P1 significantly relax the limit on the gluino mass to 450. 
This is the largest relaxation of $\mgl$ limit we have noted in this paper. 
In mSUGRA the limit L4 is obtained by the last cut set (`High Mass') employed by 
the ATLAS group \cite{atlas0l}. The efficiencies for these cuts are presented in Table 3. 
For the same average squark mass and $\mgl$, the corresponding efficiencies for 
the point P1 are displayed in Table 3 which are found to be smaller in each case. 
Models with EW sectors as in P2 (P3) relaxes (strengthens) the  gluino mass limit to 500 (620).


We next consider  the light stop scenario (i.e., $\lstop$ is the 
only strongly interacting sparticle within the reach of 7 TeV 
experiments) and examine the potential of the $blj$$\etslash$ and $b\tau j$$\etslash$
signals stemming from $\lstop$ pair production for  various EW sectors
chosen from Fig 1. For this analysis we have taken the third generation squark mass parameters 
($M_{\tilde q_3}$,$M_{\tilde u_3}$ and $M_{\tilde d_3}$) as free 
parameters. For $\lstop$ masses within the reach of the 
LHC experiments at 7 TeV, this scenario is not constrained at all by the jets + 
$\etslash$ data obtained with hard cuts on kinematic variables like
$\etslash$ and $m_{eff}$.    

\begin{table}[!htb]
\begin{center}
\begin{tabular}{|c||c||c|}
\hline
 EW sector	&S ($blj$$\etslash$)  			&S ($b\tau j$$\etslash$)	\\
 same as	&({\it Cut Set 1} of \cite{arghya1})	&({\it Cut Set 1} of \cite{arghya2})	\\
\hline
P1	 	&426(4.8*)				&82(2.0**)			\\
\hline
P2	 	&449(5.0)				&8(0.2)				\\
\hline
P3	 	&309(3.4*)				&313(7.5)			\\
\hline
P4	 	&249(2.8*)				&233(5.6)			\\
\hline
P5	 	&456(5.1)				&64(1.5)			\\
\hline
P6	 	&251(2.8*)				&2(0.04)			\\
\hline
P7	 	&219(2.4*)				&24(0.6)			\\
\hline
       \end{tabular}
       \end{center}
          \caption{ {\small Number of events and the significance for $\lum$  = 1 $\ifb$ 
from pure stop-stop production using NLO cross sections from PROSPINO. 
Significance for each case is given in the parentheses and entries marked with * (**) indicate 
that the signal is observable for 1 $<$ $\lum$ $\leq$  5 $\ifb$ (5 $<$ $\lum$ $\leq$  10 $\ifb$). 
For this table $\mlstop $ is 207. }}
          \end{table}

We have used softer cuts proposed in \cite{arghya2,arghya1}. 
The  $\mlstop$ mass reach via the above signals has been checked  using 
different EW scenarios.

For the $blj$$\etslash$, $l=e$ and $\mu$, signal we have used the following cuts 
from \cite{arghya1}. We have selected events with one isolated lepton, 
one {\it tagged} $b$ jet, at least 2 jets, $\etslash$ $\ge$ 75 and 
$P_T$ of {\it tagged} $b$ jet $\le 80$. The SM backgrounds after these cuts 
are available in Table 6 of \cite{arghya1}.

For the $b \tau j$$\etslash$ signal we have used the following cuts from 
\cite{arghya2} which studied this signal for the first time. 
We have demanded events with one {\it tagged} $b$ jet, 
one {\it tagged} $\tau $ jet, no isolated lepton, $\etslash \ge 70$ and 
$P_T$ of {\it tagged} $\tau$ jet $\ge 40$. SM background details after 
these cuts are available in Table 1 of \cite{arghya2}.

In the light stop scenario if BR ($\charginopm$ $\ra \stauone \nu_{\tau} , 
\snu_{\tau} \tau$)=100 $\%$ then it is estimated that $\mlstop \leq$ 280 (305) 
can be probed by $b\tau j$$\etslash$ signal with $\lum = 5 (10) \ifb$. This is 
similar to the $\mlstop$ reach obtained in  \cite{arghya2} where the light EW sector
was taken to be as in mSUGRA. Typical signal sizes for the benchmark points and 
the corresponding (S/$\sqrt B$) are presented in Table 4 for $\mlstop$ = 207. 
The sensitivity of the signal to the EW sector is illustrated by this table.

Models with EW sectors as in P2 and P5 (with large leptonic BR of 
the chargino) give the maximum stop mass reach with $blj$$\etslash$ signal. 
For these models $\mlstop \leq$ 275 (305) can be probed by $b l j 
\etslash$ signal with $\lum = 5 (10) \ifb$. Otherwise stop mass reach 
with $\lum = 5 (10) \ifb$ varied between 230-260 (260-290) for the rest 
of the benchmark points.

It has been noted that the $b\tau j$$\etslash$ signal will be disfavoured 
if $\stauone$ is the NLSP and $\stauone - \lspone$ co-annihilation is the main 
mechanism for relic density production. Similar degradation of the $blj$$\etslash$
signal occurs if $\chonepm$ is the NLSP and $\chonepm - \lspone$ co-annihilation 
contributes significantly to the relic density production. Both the signal will be 
degraded if $\lstop$ and $\chonepm$ masses are close together leading to difficulties in 
b tagging.

We next consider the $blj$$\etslash$ and the $b \tau j$$\etslash$ signal
in the LSG scenario discussed in the introduction following \cite{arghya2}. 
In this scenario we have treated $M_3$ as a free parameter in addition to the 
soft breaking parameters for the third generation squarks. Here all squarks except 
$\lstop$ are assumed to be beyond the reach of the LHC  7 TeV experiments. We 
use different EW sectors in the unconstrained MSSM 
from Fig. 1.  

The first information required is the lower limit on $\mgl$, if any,
in this scenario.   
As noted earlier the limits on $\mgl$ in models where 
all squarks (including $\lstop$) are assumed to be heavy and  
gluinos are light are changed significantly in the unconstrained
MSSM. The lower limits on $\mgl$ in P1, P2 and P3 are 450, 
500, 620 respectively
from  jets + $\etslash$ data.  One would naively expect the above limits
to be valid in the LSG scenario as well since the 
strong cuts of ATLAS eliminates the signal from light stop pairs if any
\cite{arghya2}. 

But a more stringent lower bound on $\mgl$ in the LSG scenario 
arises from the 
$blj$$\etslash$ 
signal \cite{atlasblj}. Using 1.03 $\ifb$ data the ATLAS group 
have excluded gluino masses below 500-520 for $\mlstop$ in the range 125 
- 300. The $\mgl$ limit is insensitive to $\mlstop$ since the hard 
ATLAS cuts eliminate the events from $\lstop$ pair production.

\begin{table}[!htb]
\begin{center}\
\begin{tabular}{|c|c|c|c|c|}
\hline
		& $t\bar t$ 	&     QCD  		& $W + 1j$    	     	& $W + 2j$  \\
\hline
$\sigma$ (pb)   &  85.5		& $7.7 \times10^7$  	& $1.4 \times10^4$  	&$5.2\times10^3$     \\
\hline
C1		 &28.728 	 &$2.2\times10^5$	&$3.1\times10^3$	&$9.53\times10^2$    \\
\hline
C2		 &14.0519	 &$1.07\times10^4$	&8.2939   		&8.5925 	\\
\hline
C3		 &3.5662	 &$1.2465\times10^2$ 	&1.1499	  		&1.9352  	 \\
\hline
C4	 	 &1.8647	 &12.9337		&0.5005	  		&0.8375   	 \\
\hline
C5	 	 &1.7219	 &0.0052		&0.4051   		&0.7097 	\\
\hline

\end{tabular}
\end{center}
\caption{ {\small The LO cross-sections (including efficiency) of the SM backgrounds after the cuts(C1 - C5) .}}
\end{table}

We shall use softer cuts \cite{arghya2,arghya1} to simulate the above signals.
However, we have optimized the cuts in \cite{arghya1} (see below). Our 
main aim is to find the stop mass reach for a fixed gluino mass 
consistent with the LHC data. For $b l j$$\etslash$ signal we have 
implemented the following optimized cuts in succession to enhance the 
signal to background ratio.
\begin{itemize}
\item We have selected events with one isolated lepton (C1).
\item We have selected events with one {\it tagged} $b$ jet (C2).
\item We have demanded events with $P_T$ of leading jet $\ge 120$ (C3).
\item Events with missing transverse energy ($\etslash) \ge 70$ are selected (C4).
\item Events with $M_{eff} \ge 300$   are selected (C5).
\end{itemize}
The response of the SM backgrounds to these cuts are 
presented in Table 5. We have also studied the $b \tau j \etslash$
signal, analysed for the first time in \cite{arghya2} using the  
cuts proposed in that paper.

For $\mgl$= 550  the reach in $\mlstop$ for $\lum$ = 10 $\ifb$
is displayed in Table 6 for different EW scenarios. 
The highest stop mass can be probed in 
P2 with our $b l j$$\etslash$ signal at $\lum$ = 10 $\ifb$ is 
375. However, the signal may also be degraded due to the degeneracies discussed in 
the light $\lstop$ scenario. Tick marks in Table 6 indicate the channel via which the 
highest $\mlstop$ can be probed. It is clear that the signals are sensitive to the EW 
sector and the two proposed signals are complements each other.

\begin{table}[!htb]
\begin{center}
\begin{tabular}{|c|c||c||c|}
\hline
 EW sector	&$ \mlstop $	& \multicolumn{2}{c|}{Reach obtained by} 			\\
\cline{3-4}
 same as	& reach		& $blj$$\etslash$ signal&$b\tau j$$\etslash$ signal	\\
\hline
P1	 	&310		&$\surd$			&-			\\
\hline
P2	 	&375		&$\surd$		&-			\\
\hline
P3	 	&330		&-			&$\surd$	\\
\hline
P4	 	&325		&-			&$\surd$		\\
\hline
P5	 	&350		&$\surd$		&-			\\
\hline
P6	 	&335		&$\surd$		&-			\\
\hline
P7	 	&305		&$\surd$		&-			\\
\hline
       \end{tabular}
       \end{center}
          \caption{ {\small Stop mass reach in different EW models for $\mgl$ = 550  for $\lum$  = 10 $\ifb$ 
in the LSG scenario using NLO cross sections. Tick mark indicates the channel for which the mass 
reach is obtainable. }}
          \end{table}


\section{Conclusions}

The present lower limits on the sparticle masses in mSUGRA obtained by 
the current experiments at the LHC are primarily governed by the 
strongly interacting sparticles (the squarks and the gluinos).

However, strong limits on sparticle masses in the EW sector-much above the 
corresponding limits from LEP- emerge due to the correlations among these 
masses in mSUGRA. For example, the mass of the LSP is constrained by 
$\mlspone \gsim 160$ \cite{boer}. This constraint may have important consequences 
for low mass neutralino DM and the direct search for the neutralinos 
\cite {akulaprannath,xenon100,boer} and neutralino mass 
reconstruction \cite{dmmass}. The importance of light neutralino scenarios of 
cosmological importance has been discussed in the context of LHC experiments by \cite {scopel}.

Since the SUSY breaking mechanism is essentially unknown till this date 
it is worthwhile to revisit the above limits in an unconstrained MSSM 
with no correlation among the sparticle masses in the strong and EW 
sectors. In order to scan the parameter space we have varied the 
sparticle masses focusing on light EW sectors as described in section 
2.1. All soft breaking masses in this sector are assumed to be below 
200. Some important departures from the mSUGRA spectrum 
which have important consequence for neutralino DM as well as new physics 
search at the LHC are S1) no correlation between the chargino 
and the LSP mass is assumed, S2) the left sleptons 
could be lighter than the right sleptons and S3) the sleptons belonging 
to the first two generations could be lighter or heavier than the 
lighter stau mass eigenstate. The present paper is, therefore, an 
extension of \cite{arghya2,arghya1} which for simplicity assumed 
the masses of the EW sparticles to be  correlated as in mSUGRA. 
The scenario S1 can be motivated by mSUGRA type model with non-universal 
gaugino masses \cite{nonunigauge} while the other two scenarios by models 
involving non-universal scalar masses \cite{polonsky,dterm}.

We then vary the parameters in the EW sector and delineate the parameter 
space consistent with the WMAP data on DM relic density (see Fig. 1). As 
a consequence of S1) neutralinos with mass $\gsim$ 35 are 
found to be consistent with WMAP data as long as the LSP is bino 
dominated or is a mixture of bino and wino. This revives the possibility 
of neutralino annihilation via a nearly on shell Z boson 
and/or lighter h-boson. Moreover a chargino NLSP is a distinct possibility 
in this scenario and chargino-neutralino co annihilation is a viable DM
producing mechanism as is readily seen from Fig 1. In S2 with relatively
light L-type sleptons and sneutrino, neutralino annihilation can proceed
efficiently. Moreover, LSP - sneutrino co annihilation may become an important
DM producing mechanism for each LSP mass. This mechanism is especially 
important for S3 when $\stauone$ is relatively heavy. 
None of these mechanisms are viable in mSUGRA in view of the LHC data.

Using the ATLAS selection criteria \cite{atlas0l} we find that in an 
unconstrained MSSM, the limits on the squarks and gluinos derived in 
mSUGRA changes either way by at most 10-12\% for most choices of the light EW 
sector consistent with the WMAP data as described in Fig. 1 (sub-section 2.3). 
This shows that i) the limits on $\msq$ and $\mgl$ derived in mSUGRA are approximately 
valid in the unconstrained MSSM and ii) EW sparticles much lighter than that in mSUGRA are consistent 
with the LHC data.  
If the leptonic BR of chargino decays are much larger 
than that in mSUGRA, then the lepton veto in the ATLAS selection procedure  
relaxes the gluino mass limits appreciably. This is illustrated by the 
points P2 and P6 which belongs to the scenario S2. 
It may be recalled that a similar relaxation was noted in the context of 
old Tevatron experiments (see the 1st paper of \cite{nonuniscapheno}). 
For very heavy squarks, the gluino mass limit L4 (see sub-section 2.3) 
reduces by 20 \% . This is the largest relaxation in $\mgl$ noted by us 
and is illustrated by the point P1 belonging to scenario S1. 
It is, therefore, fair to conclude that the current 
LHC data in the jets + $\etslash$ channel is by and large insensitive to 
the choice of the EW sector and EW sparticles having masses just above 
the corresponding LEP limits are very much allowed.

In the absence of a suitable $e^+ - e^-$ collider it not easy to 
directly test the EW sector even if the sparticles in this sector are 
light. The simulations of the LHC 14 TeV experiments \cite{cmstdr}
suggest that the clean tri-lepton signal from chargino-neutralino 
production or the di-lepton + $\etslash$ signal stemming from light 
slepton pair production may not be viable at the experiments at 7 TeV.

We have, therefore, considered two scenarios each with a subset of the strongly
interacting sparticles and a light  EW sector within the reach of the LHC,
They are i) the light stop scenario and ii) the light stop gluino (LSG)
scenario. In both cases we consider $blj\etslash$ \cite{atlasblj,arghya2} and
$b\tau j$$ \etslash$ \cite{arghya1} signal. We also follow the selection 
criteria suggested in these references.

In the light stop scenario we find that if the lighter chargino decays 
into $\tau$ rich final states with large BR (see, e.g., points P3 and P4 
in Table 2) $\mlstop$ upto 260 (305) can be probed by the $b\tau j$$\etslash$ signal, 
investigated for the first time in \cite{arghya2}, for $\lum$ = 5 (10) $\ifb$. 
On the other hand if $\charginopm$ decays into e or $\mu$ with large BR 
(see,e.g.,P2 and P5) $\mlstop$ upto 275 (305) can be probed by the $blj$$\etslash$ signal 
for $\lum$ = 5 (10) $\ifb$. For the rest of the benchmark points the
mass reach varies between 230 - 260 (260 - 290) for $\lum$ = 5 (10) $\ifb$.
It was noted in \cite{arghya2,arghya1} that the signal size is quite 
sensitive to the EW sector. This is also reflected by Table 6.
It should, however, be borne in mind that the above signals can be degraded
due to some unexpected degeneracies in the spectrum as discussed in sub-section 2.3.

 Signal of the LSG scenario in the $b l j \etslash$ channel has also 
been considered by the ATLAS collaboration. However, as pointed out in 
\cite{arghya2,arghya1} they use very hard cuts which practically 
eliminate the signal from $\lstop$ pair production. As a result the 
gluino mass reach obtained by them is practically insensitive to 
$\mlstop$. Using softer cuts following \cite{arghya2,arghya1} we find the 
reach in $\mlstop$ for a fixed $\mgl$= 550 and display the results
in Table 6. The signal size indeed show some sensitivity to the
EW sector. Of course the signal will be degraded if the degeneracies
as discussed above are present. We conclude by noting that supersymmetry 
with a light electroweak sector consistent with the WMAP and LEP constraints 
is very much allowed by the current LHC data.

\end{document}

%% file: definition.tex
\topmargin -0.1in
\headsep 30pt
\footskip 40pt
\oddsidemargin 12pt
\evensidemargin -16pt
\textheight 8.5in
\textwidth 6.5in
\parindent 20pt
 
\def\baselinestretch{1.5}
\newcommand{\newc}{\newcommand}
\def\preprint{{preprint}}
\def\Ord{\lower .7ex\hbox{$\;\stackrel{\textstyle <}{\sim}\;$}}
\def\OOrd{\lower .7ex\hbox{$\;\stackrel{\textstyle >}{\sim}\;$}}
\def\cO#1{{\cal{O}}\left(#1\right)}
\newc{\order}{{\cal O}}
\def\lag             {{\cal L}}
\def\Lag             {{\cal L}}
\def\lum             {{\cal L}}
\def\R               {{\cal R}}
\def\Rsq             {{\cal R}^{\sq}}
\def\Rst             {{\cal R}^{\st}}
\def\Rsb             {{\cal R}^{\sb}}
\def\M               {{\cal M}}
\def\Oas             {{\cal O}(\alpha_{s})}
\def\Vcal            {{\cal V}}
\def\Wcal            {{\cal W}}
\newc{\be}{\begin{equation}}
\newc{\ee}{\end{equation}}
\newc{\br}{\begin{eqnarray}}
\newc{\er}{\end{eqnarray}}
\newc{\ba}{\begin{array}}
\newc{\ea}{\end{array}}
\newc{\bi}{\begin{itemize}}
\newc{\ei}{\end{itemize}}
\newc{\bn}{\begin{enumerate}}
\newc{\en}{\end{enumerate}}
\newc{\bc}{\begin{center}}
\newc{\ec}{\end{center}}
\newc{\ul}{\underline}
\newc{\ol}{\overline}
\newc{\ra}{\rightarrow}
\newc{\lra}{\longrightarrow}
\newc{\wt}{\widetilde}
\newc{\til}{\tilde}
\def\kr              {^{\dagger}}
\newc{\wh}{\widehat}
\newc{\ti}{\times}
\newc{\Dir}{\kern -6.4pt\Big{/}}
\newc{\Dirin}{\kern -10.4pt\Big{/}\kern 4.4pt}
\newc{\DDir}{\kern -10.6pt\Big{/}}
\newc{\DGir}{\kern -6.0pt\Big{/}}
\newc{\sig}{\sigma}
\newc{\sigmalstop}{\sig_{\lstoppair}}
\newc{\Sig}{\Sigma}  
\newc{\del}{\delta}
\newc{\Del}{\Delta}
\newc{\lam}{\lambda}
\newc{\Lam}{\Lambda}
\newc{\gam}{\gamma}
\newc{\Gam}{\Gamma}
\newc{\eps}{\epsilon}
\newc{\Eps}{\Epsilon}
\newc{\kap}{\kappa}
\newc{\Kap}{\Kappa}
\newc{\modulus}[1]{\left| #1 \right|}
\newc{\eq}[1]{(\ref{eq:#1})}
\newc{\eqs}[2]{(\ref{eq:#1},\ref{eq:#2})}
\newc{\etal}{{\it et al.}\ }
\newc{\ibid}{{\it ibid}.}
\newc{\ibidem}{{\it ibidem}.}
\newc{\eg}{{\it e.g.}\ }
\newc{\ie}{{\it i.e.}\ }
\def \viz{\emph{viz.}}
\def \etc{\emph{etc. }}
\newc{\nonum}{\nonumber}
\newc{\lab}[1]{\label{eq:#1}}
\newc{\dpr}[2]{({#1}\cdot{#2})}
\newc{\lt}{\stackrel{<}}
\newc{\gt}{\stackrel{>}}
\newc{\lsimeq}{\stackrel{<}{\sim}}
\newc{\gsimeq}{\stackrel{>}{\sim}}
\def\lsim{\buildrel{\scriptscriptstyle <}\over{\scriptscriptstyle\sim}}
\def\gsim{\buildrel{\scriptscriptstyle >}\over{\scriptscriptstyle\sim}}
\def\lapp{\mathrel{\rlap{\raise.5ex\hbox{$<$}}
                    {\lower.5ex\hbox{$\sim$}}}}
\def\gapp{\mathrel{\rlap{\raise.5ex\hbox{$>$}}
                    {\lower.5ex\hbox{$\sim$}}}}
\newc{\half}{\frac{1}{2}}
\newcommand {\nnc}        {{\overline{\mathrm N}_{95}}}
\newcommand {\dm}         {\Delta m}
\newcommand {\dM}         {\Delta M}
\def\bra{\langle}
\def\ket{\rangle}
\def\cO#1{{\cal{O}}\left(#1\right)}
\def \DM{{\Delta{m}}}
\newc{\bQ}{\ol{Q}}
\newc{\dota}{\dot{\alpha }}
\newc{\dotb}{\dot{\beta }}
\newc{\dotd}{\dot{\delta }}
\newc{\nindnt}{\noindent}

\newcommand{\medf}[2] {{\footnotesize{\frac{#1}{#2}} }}
\newcommand{\smaf}[2] {{\textstyle \frac{#1}{#2} }}
\def\onesq            {{\textstyle \frac{1}{\sqrt{2}} }}
\def\onehf            {{\textstyle \frac{1}{2} }}
\def\oneth            {{\textstyle \frac{1}{3} }}
\def\twoth            {{\textstyle \frac{2}{3} }}
\def\onefo            {{\textstyle \frac{1}{4} }}
\def\forth            {{\textstyle \frac{4}{3} }}

\newc{\matth}{\mathsurround=0pt}
\def\ML{\ifmmode{{\mathaccent"7E M}_L}
             \else{${\mathaccent"7E M}_L$}\fi}
\def\MR{\ifmmode{{\mathaccent"7E M}_R}
             \else{${\mathaccent"7E M}_R$}\fi}
\newcommand{\s}{\\ \vspace*{-3mm} }

\def \ud { {1 \over 2} }
\def \ut { {1 \over 3} }
\def \td { {3 \over 2} }
\newc{\mr}{\mathrm}
\def\dh {\partial }
\def \cs { cross-section }
\def \css { cross-sections }
\def \cm { centre of mass }
\def \cms { centre of mass energy }
\def \cc { coupling constant }
\def \ccs {coupling constants }
\def \gc {gauge coupling }
\def \gcc {gauge coupling constant }
\def \gccs {gauge coupling constants }
\def \yc {Yukawa coupling }
\def \ycc {Yukawa coupling constant }
\def \pp {{parameter }}
\def \pps {{parameters }} 
\def \ps {parameter space }
\def \pss {parameter spaces }
\def \vv {vice versa }

\newc{\siminf}{\mbox{$_{\sim}$ {\small {\hspace{-1.em}{$<$}}}    }}
\newc{\simsup}{\mbox{$_{\sim}$ {\small {\hspace{-1.em}{$>$}}}    }}


\newc {\Zboson}{{\mathrm Z}^{0}}
\newc{\thetaw}{\theta_W}
\newc{\mbot}{{m_b}}
\newc{\mtop}{{m_t}}
\newc{\sm}{${\cal {SM}}$}
\newc{\as}{\alpha_s}
\newc{\aem}{\alpha_{em}}
\def \PI{{\pi^{\pm}}}
\newc{\ppbar}{\mbox{$p\ol{p}$}}
\newc{\bbbar}{\mbox{$b\ol{b}$}}
\newc{\ccbar}{\mbox{$c\ol{c}$}}
\newc{\ttbar}{\mbox{$t\ol{t}$}}
\newc{\eebar}{\mbox{$e\ol{e}$}}
\newc{\zzero}{\mbox{$Z^0$}}
\def \gamz{\Gam_Z}
\newc{\wplus}{\mbox{$W^+$}}
\newc{\wminus}{\mbox{$W^-$}}
\newc{\ellp}{\ell^+}
\newc{\ellm}{\ell^-}
\newc{\elp}{\mbox{$e^+$}}
\newc{\elm}{\mbox{$e^-$}}
\newc{\elpm}{\mbox{$e^{\pm}$}}
\newc{\qbar}     {\mbox{$\ol{q}$}}
\def \ewgroup{SU(2)_L \otimes U(1)_Y}
\def \smgroup{SU(3)_C \otimes SU(2)_L \otimes U(1)_Y}
\def \smcolorem{SU(3)_C \otimes U(1)_{em}}

\def \SSM  {Supersymmetric Standard Model}
\def \poincare{Poincare$\acute{e}$}
\def \superspace{\emph{superspace}}
\def \sfs{\emph{superfields}}
\def \superpot{\emph{superpotential}}
\def \csf{\emph{chiral superfield}}
\def \csfs{\emph{chiral superfields}}
\def \vsf{\emph{vector superfield }}
\def \vsfs{\emph{vector superfields}}
\newc{\Ebar}{{\bar E}}
\newc{\Dbar}{{\bar D}}
\newc{\Ubar}{{\bar U}}
\newc{\susy}{{{SUSY}}}
\newc{\msusy}{{{M_{SUSY}}}}

\def\photino{\ifmmode{\mathaccent"7E \gam}\else{$\mathaccent"7E \gam$}\fi}
\def\taugluino{\ifmmode{\tau_{\mathaccent"7E g}}
             \else{$\tau_{\mathaccent"7E g}$}\fi}
\def\mphotino{\ifmmode{m_{\mathaccent"7E \gam}}
             \else{$m_{\mathaccent"7E \gam}$}\fi}
\newc{\gl}   {\mbox{$\wt{g}$}}
\newc{\mgl}  {\mbox{$m_{\gl}$}}
\def \charginopm{{\wt\chi}^{\pm}}
\def \mcharginopm{m_{\charginopm}}
\def \mchpmmin {\mcharginopm^{min}}
\def \chonep {{\wt\chi_1^+}}
\def \chone {{\wt\chi_1}}
\def \ch2p {{\wt\chi_2^+}}
\def \chonem {{\wt\chi_1^-}}
\def \ch2m {{\wt\chi_2^-}}
\def \chplus {{\wt\chi^+}}
\def \chminus {{\wt\chi^-}}
\def \chonip{{\wt\chi_i}^{+}}
\def \chonim{{\wt\chi_i}^{-}}
\def \chonipm{{\wt\chi_i}^{\pm}}
\def \chonjp{{\wt\chi_j}^{+}}
\def \chonjm{{\wt\chi_j}^{-}}
\def \chonjpm{{\wt\chi_j}^{\pm}}
\def \chonepm{{\wt\chi_1}^{\pm}}
\def \chonemp{{\wt\chi_1}^{\mp}}
\def \mchonepm{m_{\chonepm}}
\def \mchonemp{m_{\chonemp}}
\def \chtwopm{{\wt\chi_2}^{\pm}}
\def \mchtwopm{m_{\chtwopm}}
\newc{\dmchi}{\Delta m_{\wt\chi}}


\def \vlsp{\emph{VLSP}}
\def \lspi{\wt\chi_i^0}
\def \mlspi{m_{\lspi}}
\def \lspj{\wt\chi_j^0}
\def \mlspj{m_{\lspj}}
\def \lspone{\wt\chi_1^0}
\def \mlspone{m_{\lspone}}
\def \lsptwo{\wt\chi_2^0}
\def \mlsptwo{m_{\lsptwo}}
\def \lspthree{\wt\chi_3^0}
\def \mlspthree{m_{\lspthree}}
\def \lspfour{\wt\chi_4^0}
\def \mlspfour{m_{\lspfour}}


\newc{\sele}{\wt{\mathrm e}}
\newc{\sell}{\wt{\ell}}
\def \msell{m_{\sell}}
\def \slepone{\wt\ell_1}
\def \mslepone{m_{\slepone}}
\def \smuone{\wt\mu_1}
\def \msmuone{m_{\smuone}}
\def \stauone{\wt\tau_1}
\def \mstauone{m_{\stauone}}
\def \snu{\wt{\nu}}
\def \snutau{\wt{\nu}_{\tau}}
\def \msnu{m_{\snu}}
\def \msnumu{m_{\snu_{\mu}}}
\def \barsnu{\wt{\bar{\nu}}}
\def \barsnul{\barsnu_{\ell}}
\def \snul{\snu_{\ell}}
\def \mbarsnu{m_{\barsnu}}
\newc{\snue}     {\mbox{$ \wt{\nu_e}$}}
\newc{\smu}{\wt{\mu}}
\newc{\stau}{\wt{\tau}}
\newc {\nuL} {\wt{\nu}_L}
\newc {\nuR} {\wt{\nu}_R}
\newc {\snub} {\bar{\wt{\nu}}}
\newc {\eL} {\wt{e}_L}
\newc {\eR} {\wt{e}_R}
\def \slepl{\wt{l}_L}
\def \mslepl{m_{\slepl}}
\def \slepr{\wt{l}_R}
\def \mslepr{m_{\slepr}}
\def \stau{\wt\tau}
\def \mstau{m_{\stau}}
\def \slepton{\wt\ell}
\def \mslepton{m_{\slepton}}
\def \mlhiggs{m_{h^0}}

\def \xr{X_{r}}

\def \sfer{\wt{f}}
\def \msfer{m_{\sfer}}
\def \sq{\wt{q}}
\def \msq{m_{\sq}}
\def \msquleft{m_{\tilde{u_L}}}
\def \msqurht{m_{\tilde{u_R}}}
\def \sql{\wt{q}_L}
\def \msql{m_{\sql}}
\def \sqr{\wt{q}_R}
\def \msqr{m_{\sqr}}
\newc{\msqot}  {\mbox{$m_(\sq_{1,2} )$}}
\newc{\sqbar}    {\mbox{$\bar{\wt{q}}$}}
\newc{\ssb}      {\mbox{$\squark\ol{\squark}$}}
\newc {\qL} {\wt{q}_L}
\newc {\qR} {\wt{q}_R}
\newc {\uL} {\wt{u}_L}
\newc {\uR} {\wt{u}_R}
\def \ul{\wt{u}_L}
\def \mul{m_{\ul}}
\newc {\dL} {\wt{d}_L}
\newc {\dR} {\wt{d}_R}
\newc {\cL} {\wt{c}_L}
\newc {\cR} {\wt{c}_R}
\newc {\sL} {\wt{s}_L}
\newc {\sR} {\wt{s}_R}
\newc {\tL} {\wt{t}_L}
\newc {\tR} {\wt{t}_R}
\newc {\stb} {\ol{\wt{t}}_1}
\newc {\sbot} {\wt{b}_1}
\newc {\msbot} {m_{\sbot}}
\newc {\sbotb} {\ol{\wt{b}}_1}
\newc {\bL} {\wt{b}_L}
\newc {\bR} {\wt{b}_R}
\def \mul{m_{\wt{u}_L}}
\def \mur{m_{\wt{u}_R}}
\def \mdl{m_{\wt{d}_L}}
\def \mdr{m_{\wt{d}_R}}
\def \mcl{m_{\wt{c}_L}}
\def \charml{\wt{c}_L}
\def \mcr{m_{\wt{c}_R}}
\newc{\csquark}  {\mbox{$\wt{c}$}}
\newc{\csquarkl} {\mbox{$\wt{c}_L$}}
\newc{\mcsl}     {\mbox{$m(\csquarkl)$}}
\def \msl{m_{\wt{s}_L}}
\def \msr{m_{\wt{s}_R}}
\def \mbl{m_{\wt{b}_L}}
\def \mbr{m_{\wt{b}_R}}
\def \mtl{m_{\wt{t}_L}}
\def \mtr{m_{\wt{t}_R}}
\def \st{\wt{t}}
\def \mst{m_{\st}}
\newc {\stopl}         {\wt{\mathrm{t}}_{\mathrm L}}
\newc {\stopr}         {\wt{\mathrm{t}}_{\mathrm R}}
\newc {\stoppair}      {\wt{\mathrm{t}}_{1}
\bar{\wt{\mathrm{t}}}_{1}}
\def \lstop{\wt{t}_{1}}
\def \lstopbar{\lstop^*}
\def \hstop{\wt{t}_{2}}
\def \hstopbar{\hstop^*}
\def \mlstop{m_{\lstop}}
\def \mhstop{m_{\hstop}}
\def \lstoppair{\lstop\lstop^*}
\def \hstoppair{\hstop\hstop^*}
\newc{\tsquark}  {\mbox{$\wt{t}$}}
\newc{\ttb}      {\mbox{$\tsquark\ol{\tsquark}$}}
\newc{\ttbone}   {\mbox{$\tsquark_1\ol{\tsquark}_1$}}
\def \tsq {top squark }
\def \tsqs {top squarks }
\def \tsql {ligtest top squark }
\def \tsqh {heaviest top squark }
\newc{\mix}{\theta_{\wt t}}
\newc{\cost}{\cos{\theta_{\wt t}}}
\newc{\sint}{\sin{\theta_{\wt t}}}
\newc{\costloop}{\cos{\theta_{\wt t_{loop}}}}
\def \lsbot{\wt{b}_{1}}
\def \lsbotbar{\lsbot^*}
\def \hsbot{\wt{b}_{2}}
\def \hsbotbar{\hsbot^*}
\def \mlsbot{m_{\lsbot}}
\def \mhsbot{m_{\hsbot}}
\def \lsbotpair{\lsbot\lsbot^*}
\def \hsbotpair{\hsbot\hsbot^*}
\newc{\mixsbot}{\theta_{\wt b}}

\def \mhone{m_{h_1}}
\def \hup{{H_u}}
\def \hdn{{H_d}}
\newc{\tb}{\tan\beta}
\newc{\cb}{\cot\beta}
\newc{\vev}[1]{{\left\langle #1\right\rangle}}

\def \abot{A_{b}}
\def \atop{A_{t}}
\def \atau{A_{\tau}}
\newc{\mhalf}{m_{1/2}}
\newc{\mzero} {\mbox{$m_0$}}
\newc{\azero} {\mbox{$A_0$}}

\newc{\lb}{\lam}
\newc{\lbp}{\lam^{\prime}}
\newc{\lbpp}{\lam^{\prime\prime}}
\newc{\rpv}{{\not \!\! R_p}}
\newc{\rpvm}{{\not  R_p}}
\newc{\rp}{R_{p}}
\newc{\rpmssm}{{RPC MSSM}}
\newc{\rpvmssm}{{RPV MSSM}}


\newc{\sbyb}{S/$\sqrt B$}
\newc{\pelp}{\mbox{$e^+$}}
\newc{\pelm}{\mbox{$e^-$}}
\newc{\pelpm}{\mbox{$e^{\pm}$}}
\newc{\epem}{\mbox{$e^+e^-$}}
\newc{\lplm}{\mbox{$\ell^+\ell^-$}}
\def \branch{\emph{BR}}
\def \branche{\branch(\lstop\ra be^{+}\nu_e \lspone)\ti \branch(\lstop^{*}\ra \bar{b}q\bar{q^{\prime}}\lspone)}
\def \branchmu{\branch(\lstop\ra b\mu^{+}\nu_{\mu} \lspone)\ti \branch(\lstop^{*}\ra \bar{b}q\bar{q^{\prime}}\lspone)}
\def\Ecm{\ifmmode{E_{\mathrm{cm}}}\else{$E_{\mathrm{cm}}$}\fi}
\newc{\rts}{\sqrt{s}}
\newc{\rtshat}{\sqrt{\hat s}}
\newc{\gev}{\,GeV}
\newc{\mev}{~{\rm MeV}}
\newc{\tev}  {\mbox{$\;{\rm TeV}$}}
\newc{\gevc} {\mbox{$\;{\rm GeV}/c$}}
\newc{\gevcc}{\mbox{$\;{\rm GeV}/c^2$}}
\newc{\intlum}{\mbox{${ \int {\cal L} \; dt}$}}
\newc{\call}{{\cal L}}
\def \met  {\mbox{${E\!\!\!\!/_T}$}}
\def \cpv  {\mbox{${CP\!\!\!\!/}$}}
\newc{\ptmiss}{/ \hskip-7pt p_T}
\def \eslash{\not \! E}
\def \etslash{\not \! E_T }
\def \ptslash{\not \! p_T }
\newc{\PT}{\mbox{$p_T$}}
\newc{\ET}{\mbox{$E_T$}}
\newc{\dedx}{\mbox{${\rm d}E/{\rm d}x$}}
\newc{\ifb}{\mbox{${\rm fb}^{-1}$}}
\newc{\ipb}{\mbox{${\rm pb}^{-1}$}}
\newc{\pb}{~{\rm pb}}
\newc{\fb}{~{\rm fb}}
\newc{\ycut}{y_{\mathrm{cut}}}
\newc{\chis}{\mbox{$\chi^{2}$}}
\def \hadron{\emph{hadron}}
\def \nlc{\emph{NLC }}
\def \lhc{\emph{LHC }}
\def \cdf{\emph{CDF }}
\def\dzero{\emptyset}
\def \tevatron{\emph{Tevatron }}
\def \lep{\emph{LEP }}
\def \jets{\emph{jets }}
\def \jet(s){\emph{jet(s) }}

\def\Crs{stroke [] 0 setdash exch hpt sub exch vpt add hpt2 vpt2 neg V currentpoint stroke 
hpt2 neg 0 R hpt2 vpt2 V stroke}
\def\loopdk{\lstop \ra c \lspone}
\def\brloopdk{\branch(\loopdk)}
\def\fourdk{\lstop \ra b \lspone  f \bar f'}
\def\brfourdk{\branch(\fourdk)}
\def\fourdklep{\lstop \ra b \lspone  \ell \nu_{\ell}}
\def\fourdkhad{\lstop \ra b \lspone  q \bar q'}
\def\brfourdklep{\branch(\fourdklep)}
\def\brfourdkhad{\branch(\fourdkhad)}
\def\twodk{\lstop \ra b \chonep}
\def\brtwodk{\branch(\twodk)}
\def\threedkslep{\lstop \ra b \wt{\ell} \nu_{\ell}}
\def\brthreedkslep{\branch(\threedkslep)}
\def\threedksnu{\lstop \ra b \wt{\nu_{\ell}} \ell }
\def\brthreedksnu{\branch(\threedksnu) }
\def\threedklsp{\lstop \ra b W \lspone }
\def\brthreedklsp{\\branch(\threedklsp) }
\def\topdk{t \ra \lstop \lspone}
\def\rpvdk{\lstop \ra e^+ d}
\def\brrpvdk{\branch(\rpvdk)}
\def\fonec{f_{11c}} 
\newc{\mpl}{M_{\rm Pl}}
\newc{\mgut}{M_{GUT}}
\newc{\mw}{M_{W}}
\newc{\mweak}{M_{weak}}
\newc{\mz}{M_{Z}}

\newc{\OPALColl}   {OPAL Collaboration}
\newc{\ALEPHColl}  {ALEPH Collaboration}
\newc{\DELPHIColl} {DELPHI Collaboration}
\newc{\XLColl}     {L3 Collaboration}
\newc{\JADEColl}   {JADE Collaboration}
\newc{\CDFColl}    {CDF Collaboration}
\newc{\DXColl}     {D0 Collaboration}
\newc{\HXColl}     {H1 Collaboration}
\newc{\ZEUSColl}   {ZEUS Collaboration}
\newc{\LEPColl}    {LEP Collaboration}
\newc{\ATLASColl}  {ATLAS Collaboration}
\newc{\CMSColl}    {CMS Collaboration}
\newc{\UAColl}    {UA Collaboration}
\newc{\KAMLANDColl}{KamLAND Collaboration}
\newc{\IMBColl}    {IMB Collaboration}
\newc{\KAMIOColl}  {Kamiokande Collaboration}
\newc{\SKAMIOColl} {Super-Kamiokande Collaboration}
\newc{\SUDANTColl} {Soudan-2 Collaboration}
\newc{\MACROColl}  {MACRO Collaboration}
\newc{\GALLEXColl} {GALLEX Collaboration}
\newc{\GNOColl}    {GNO Collaboration}
\newc{\SAGEColl}  {SAGE Collaboration}
\newc{\SNOColl}  {SNO Collaboration}
\newc{\CHOOZColl}  {CHOOZ Collaboration}
\newc{\PDGColl}  {Particle Data Group Collaboration}

\def\issue(#1,#2,#3){{\bf #1}, #2 (#3)}
\def\ASTR(#1,#2,#3){Astropart.\ Phys. \issue(#1,#2,#3)}
\def\AJ(#1,#2,#3){Astrophysical.\ Jour. \issue(#1,#2,#3)}
\def\AJS(#1,#2,#3){Astrophys.\ J.\ Suppl. \issue(#1,#2,#3)}
\def\APP(#1,#2,#3){Acta.\ Phys.\ Pol. \issue(#1,#2,#3)}
\def\JCAP(#1,#2,#3){Journal\ XX. \issue(#1,#2,#3)} 
\def\SC(#1,#2,#3){Science \issue(#1,#2,#3)}
\def\PRD(#1,#2,#3){Phys.\ Rev.\ D \issue(#1,#2,#3)}
\def\PR(#1,#2,#3){Phys.\ Rev.\ \issue(#1,#2,#3)} 
\def\PRC(#1,#2,#3){Phys.\ Rev.\ C \issue(#1,#2,#3)}
\def\NPB(#1,#2,#3){Nucl.\ Phys.\ B \issue(#1,#2,#3)}
\def\NPPS(#1,#2,#3){Nucl.\ Phys.\ Proc. \ Suppl \issue(#1,#2,#3)}
\def\NJP(#1,#2,#3){New.\ J.\ Phys. \issue(#1,#2,#3)}
\def\JP(#1,#2,#3){J.\ Phys.\issue(#1,#2,#3)}
\def\PL(#1,#2,#3){Phys.\ Lett. \issue(#1,#2,#3)}
\def\PLB(#1,#2,#3){Phys.\ Lett.\ B  \issue(#1,#2,#3)}
\def\ZP(#1,#2,#3){Z.\ Phys. \issue(#1,#2,#3)}
\def\ZPC(#1,#2,#3){Z.\ Phys.\ C  \issue(#1,#2,#3)}
\def\PREP(#1,#2,#3){Phys.\ Rep. \issue(#1,#2,#3)}
\def\PRL(#1,#2,#3){Phys.\ Rev.\ Lett. \issue(#1,#2,#3)}
\def\MPL(#1,#2,#3){Mod.\ Phys.\ Lett. \issue(#1,#2,#3)}
\def\RMP(#1,#2,#3){Rev.\ Mod.\ Phys. \issue(#1,#2,#3)}
\def\SJNP(#1,#2,#3){Sov.\ J.\ Nucl.\ Phys. \issue(#1,#2,#3)}
\def\CPC(#1,#2,#3){Comp.\ Phys.\ Comm. \issue(#1,#2,#3)}
\def\IJMPA(#1,#2,#3){Int.\ J.\ Mod. \ Phys.\ A \issue(#1,#2,#3)}
\def\MPLA(#1,#2,#3){Mod.\ Phys.\ Lett.\ A \issue(#1,#2,#3)}
\def\PTP(#1,#2,#3){Prog.\ Theor.\ Phys. \issue(#1,#2,#3)}
\def\RMP(#1,#2,#3){Rev.\ Mod.\ Phys. \issue(#1,#2,#3)}
\def\NIMA(#1,#2,#3){Nucl.\ Instrum.\ Methods \ A \issue(#1,#2,#3)}
\def\JHEP(#1,#2,#3){J.\ High\ Energy\ Phys. \issue(#1,#2,#3)}
\def\EPJC(#1,#2,#3){Eur.\ Phys.\ J.\ C \issue(#1,#2,#3)}
\def\RPP (#1,#2,#3){Rept.\ Prog.\ Phys. \issue(#1,#2,#3)}
\def\PPNP(#1,#2,#3){ Prog.\ Part.\ Nucl.\ Phys. \issue(#1,#2,#3)}
\newc{\PRDR}[3]{{Phys. Rev. D} {\bf #1}, Rapid  Communications, #2 (#3)}